\documentclass[useAMS,usenatbib]{mn2e}
\usepackage{graphicx}
\usepackage{amsmath}
\usepackage{amssymb}
\usepackage[hyphens]{url}

\def\sne{SNe Ia}
\def\sn{SN Ia}
\def\bq{\begin{quotation}}
\def\eq{\end{quotation}}
\def\be{\begin{eqnarray}}
\def\ee{\end{eqnarray}}
\def\Om{$\Omega_{\rmn{m}0}$}
\def\d{\rmn{d}}
\def\lcdm{$\Lambda$CDM}
\def\PRL#1{Phys.\ Rev.\ Lett.\ {\bf#1}} \def\PR#1{Phys.\ Rev.\ {\bf#1}}
\def\ApJ#1{ApJ {\bf#1}} \def\AJ#1{AJ {\bf#1}}

\def\MNRAS#1{MNRAS {\bf#1}}
\def\CQG#1{Class.\ Quantum Grav.\ {\bf#1}}
\def\GRG#1{Gen.\ Relativ.\ Grav.\ {\bf#1}}
\def\lsim{\mathop{\hbox{${\lower3.8pt\hbox{$<$}}\atop{\raise0.2pt\hbox{$\sim$}}$}}}
 
\def\Z#1{_{\lower2pt\hbox{$\scriptstyle#1$}}}
\def\X#1{_{\lower2pt\hbox{$\scriptstyle#1$}}}
\def\w#1{\,\hbox{#1}} \def\kmsmpc{\w{km}\;\w{sec}^{-1}\w{Mpc}^{-1}}

\def\OM{\Omega} \def\bOM{\bar\OM} \def\OmMn{\OM_{\rmn{m}}}
\def\finfty{{\mathop{\hbox{\it fi}}}}

\def\fvn{f_{\rmn{v}0}}   
\def\OMM{\bOM_{\rmn{m}0}}\def\OMk{\bOM_{\rmn{k}0}}

\def\Eiso{E_{\rmn{iso}}} \def\Epi{E_{\rmn{p,i}}}
\def\lagt{\tau_{\rmn{lag}}} \def\Epeak{E_{\rmn{peak}}}
\def\tjet{t_{\rmn{jet}}}
\def\Egamma{E_{\gamma}} \def\trt{\tau_{\rmn{RT}}}
\def\Pbolo{P_{\rmn{bolo}}} \def\Sbolo{S_{\rmn{bolo}}}
\def\thetajet{\theta_{\rmn{jet}}} \def\Fbeam{F_{\rmn{beam}}}
\def\chisq{\chi^2}

\title[Gamma ray burst distances and the timescape cosmology]{Gamma ray burst
distances and the timescape cosmology}
\author[Smale]{Peter R. Smale\thanks{E-mail:
peter.smale@pg.canterbury.ac.nz} \\
Department of Physics \& Astronomy, University of Canterbury,
Private Bag 4800, Christchurch 8140, New Zealand\\
}
\date{\today}
\begin{document}
\maketitle
\label{firstpage}
\begin{abstract}
Gamma ray bursts can potentially be used as distance indicators, providing the
possibility of extending the Hubble diagram to redshifts $\sim7$. Here we follow
the analysis of \citet{s07}, with the aim of distinguishing the timescape
cosmological model from the \lcdm\ model by means of the additional leverage
provided by GRBs in the range $2\lesssim z\lesssim7$. We find that the timescape
model fits the GRB sample slightly better than the \lcdm\ model, but that the
systematic uncertainties are still too little understood to distinguish the
models.

\end{abstract}

\section{Introduction}
The timescape (TS) model is an inhomogeneous cosmological model that explains
the apparently accelerated cosmic expansion first observed in the supernova
luminosity distances as an artifact of gradients in gravitational energy between
gravitationally bound systems and the intervening negatively curved voids. In
the dynamical spacetime of general relativity, this leads to a variance in the
calibration of the clock rates of ideal observers who fit average smoothed-out
geometries to the underlying inhomogeneous matter distribution. The TS model
agrees closely with the \lcdm\ model over the range of scales probed by the
supernova data~\citep{LNW}, with certain qualifications: parameter values
obtained by minimizing $\chisq$ fits to the TS Hubble curve depend significantly
on the process used to reduce the \sn\ light curves~\citep{sw}. 

\begin{figure}
\begin{center}
\caption{Effective comoving distance as a function of redshift for various
spatially flat models (dotted lines) and for the TS model with $\fvn=0.762$
(solid line). Parameter values for the dotted lines are (i) $\OmMn=0.249$
(best-fit to WMAP only); (ii) $\OmMn=0.279$ (joint best-fit to WMAP, BAO and
\sne); (iii) $\OmMn=0.34$ (best-fit to \citet{gold} \sne\ only). After
\citet{obs}. \label{fig:hod}}
\includegraphics[scale=0.4]{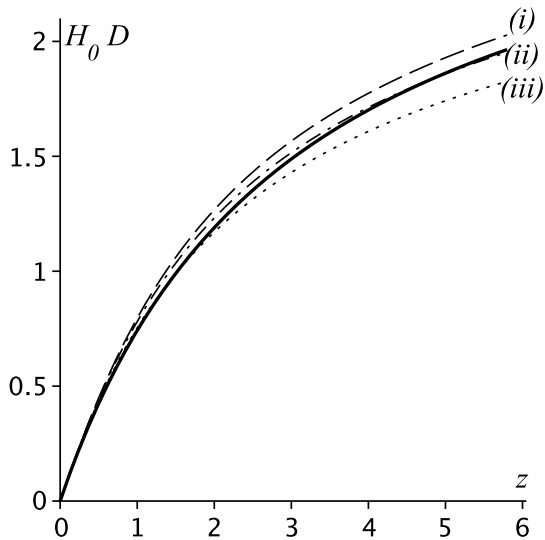}
\end{center}
\end{figure}

The current state of knowledge of systematic uncertainties in the \sn\ data 
precludes discrimination between the TS and \lcdm\ models using \sne\
\citep{sw}. In fact, calculation of the effective comoving distance $H_0D(z)$
shows that in the redshift range probed by \sne\ there is little to distinguish
between the TS model with the best-fit value for the present void fraction
$\fvn=0.762$ from the Gold dataset of \citet{gold}. \citet{obs} has noted that
over different redshift ranges $H_0D(z)$ for the TS model closely approximates
$H_0D(z)$ for spatially flat \lcdm\ models with different values of \Om\ and
$\Omega_{\Lambda0}$. It is thus seen to interpolate between different \lcdm\
models as the redshift is varied (see Fig.~\ref{fig:hod}). Fig.~\ref{fig:hod}
shows that between $z\simeq2$ and $z\simeq6$, the TS $H_0D(z)$ crosses from
coinciding closely with the best fit line from the \sne\ only to that predicted
by the best fit to WMAP, BAO and the \sne. In principle, Gamma Ray Bursts
(GRBs), which probe this redshift range, could distinguish the TS and \lcdm\
models in this redshift range, although their use as distance indicators is far
from established.

This paper will establish that the TS model is also supported by the current GRB
data~\citep{s07}, but that, as one might expect from the \sne\ results, the 
uncertainties in the data are as yet too large to distinguish the models in
the redshift range $2<z<6$.

The paper is organized as follows. Section~\ref{background} explains the method
of ``standardizing'' the GRBs for their use as distance indicators, and gives a
brief derivation of the TS luminosity distance. Section~\ref{results} describes
the results, before a discussion and conclusion are presented in
Section~\ref{discussion}.

\section[]{Background}\label{background}
\subsection[]{The timescape model}
In keeping with current observations that the large-scale cosmic structure
consists of voids of average diameter $\sim30~h^{-1}$
Mpc~\citep{hoyle2004,pan2011} separated and threaded by walls and filaments
containing clusters of galaxies, the timescape model is based on a
differentiation of the Universe into gravitationally bound spatially flat wall
regions and negatively curved voids. There is a consequent small backreaction
($\leq 5\%$ as a normalized energy density) which nevertheless leads to
significant cosmological effects over cosmological timescales~\citep{clocks}. 
At late epochs the construction of a single smoothed-out geometry becomes
problematic when the underlying geometry varies. The TS model is based on the
assumption that different equivalent descriptions of a smoothed-out average
geometry can be given, but these descriptions will vary between canonical
observers who each assume that the average geometry has the same spatial
curvature as the locally determined geometry. Differences in the calibration of
rulers and clocks grow cumulatively as the variance in spatial geometry grows,
and these must be taken into account when reconstructing the expansion history
of the universe from information on null geodesics.

As observers in galaxies, our local average geometry, assumed to be spatially
flat with scale factor $a_{\rmn{w}}$, is given by
\be \label{eq:wallmetric}
\d s^2_{\finfty}=-\d\tau^2+a_{\rmn{w}}^2(\tau)[\d\eta_{\rmn{w}}^2 +
\eta_{\rmn{w}}^2\d\Omega^2].
\ee
Finite infinity~\citep{fit1}, denoted by $\finfty$, demarcates the boundary between
gravitationally bound and unbound systems~\citep{clocks}. A similar expression 
defines the negatively curved geometry at the centre of a void, with an
appropriate void time parameter $\tau_{\rmn{v}}$ and scale factor
$a_{\rmn{v}}$. 

The volume averaged scale factor \be\label{volumeaveragescalefactor}\bar{a}^3 =
f_{\rmn{v}_i}a_{\rmn{v}}^3 + (1-f_{\rmn{v}_i})a_{\rmn{w}}^3,\ee where
$f_{\rmn{v}_i}\ll1$ is the initial void
fraction, evolves according to an averaging of the Einstein equations for
an inhomogeneous dust cosmology due to \citet{Buchert2000}. The corresponding
averaged geometry, in terms of the proper time $t$ at a volume average position
in freely expanding space, has the form
\be \label{eq:baremetric}
\d s^2=-\d t^2+\bar{a}^2(t)\d\bar{\eta}^2 +
A(\bar{\eta},t)\d\Omega^2, 
\ee
where the area function $A$ is defined by an average over the particle horizon
volume~\citep{clocks}. The local time in a wall is related to $t$ by the
phenomenological lapse parameter $\bar{\gamma}=\frac{\rmn{d}t}{\rmn{d}\tau}$.

A single geometry that relates metrics~(\ref{eq:wallmetric}) and
(\ref{eq:baremetric}) is constructed by matching the radial null geodesics
of wall and volume average geometries sharing a common centre~(\citet{clocks},
$\S$~5.2). Along the radial null geodesics, the line elements of the two
geometries are simply related by a conformal factor. Once the metric is extended
to cosmological scales, instead of (\ref{eq:baremetric}) wall observers
describe the large-scale universe by the effective metric
\be \label{eq:dressedmetric}
\d s^2=-\d\tau^2+\frac{\bar{a}^2}{\bar{\gamma^2}}[\d\bar{\eta}^2 +
r_{\rmn{w}}^2(\bar{\eta},t)\d\Omega^2],
\ee
where
$r_{\rmn{w}}\equiv\bar{\gamma}(1-f_{\rmn{v}})^{1/3}(1-f_{\rmn{v}_i})^{-1/3}\eta_
{\rmn{w}}(\bar{\eta},t)$. Metric~(\ref{eq:dressedmetric}) is
the ``dressed'' geometry that arises when we attempt to reconcile our position
as observer within a finite infinity region with observations of objects at 
cosmological distances. In general, observers (localised within finite
infinity regions) cannot assume that their measurements of cosmological
parameters correspond to the global average values. However, due to the
existence of a scale of statistical homogeneity, the values of locally measured
cosmological parameters should converge towards their global average values as
the averaging volume increases.

Volume average (``bare'') parameters, referred to metric~(\ref{eq:baremetric}),
differ from dressed parameters. The dressed matter density is
$\OmMn=\bar{\gamma}^3\OMM$, and the Hubble parameter of
metric~(\ref{eq:dressedmetric}) is related to the bare Hubble parameter
according to
\be \label{eq:hubble}
H=\bar{\gamma}\bar{H}-\frac{\d}{\d t}\bar{\gamma}. 
\ee
It is a feature of the TS model that the variance of parameters
calculated from observations of nearby objects (on scales $\lesssim
100~h^{-1}~$Mpc) will be relatively large because such objects lie within the
scale of statistical homogeneity. Since the volume of space is dominated by
voids --- of typical diameter $\sim30~h^{-1}~$Mpc --- which appear to expand
faster than the walls, observers in a typical galaxy looking on a typical line
of sight through local voids will infer that their local universe is expanding
faster than the global average. Eventually a typical line of sight will
intersect a sufficient number of walls as well as voids to approach the global
average, which does not change by sampling on ever larger scales. The
transition scale from from large to small variance in the expansion must be
larger than the diameter of the dominant voids, and is referred to as the
\emph{scale of statistical homogeneity}. It is expected to be comparable to the
BAO scale, $\sim 100~h^{-1}~$Mpc \citep{equiv}. Thus any observer in a galaxy
will typically see a ``Hubble bubble'' on scales $\lesssim 100~h^{-1}~$Mpc.

In terms of volume average time, $t$, the luminosity distance for wall observers
is~\citep{sol}
\be \label{eq:tslumdist1}
d_L =\bar{a}_0(1+z)r_{\rmn{w}}.
\ee 
The Buchert equations have an exact general solution which admits a particular
late-time attractor solution, to which the general solution converges to 
within $1\%$ by redshift $z\sim37$~\citep{sol}. For this tracker solution,
eq.~(\ref{eq:tslumdist1}) becomes~\citep{obs}
\be \label{eq:tslumdist}
\bar{H}_0d_L&=&(1+z)^2(\bar{H}_0t)^{2/3}\int_{t}^{t_0}\frac{2\bar{H}_0\rmn{d}t}{
(2+f_ { \rmn { v } } (t '))(\bar{H}_0t)^{2/3}}\\\nonumber
&=&(1+z)^2y^2\times\\\nonumber
& &\hspace{-0.5cm}\Big[2y+\frac{b}{6}\ln\Big(\frac{(y+b)^2}{y^2-by+b^2}\Big)+\frac{b
} { \sqrt { 3 } } \tan^ { -1}\Big(\frac{2y-b}{\sqrt{3}b}\Big)\Big]^{y_0}_y
\ee
where $y^3\equiv\bar{H}_0t$ and
$b^3\equiv2(1-f_{\rmn{v}0})(2+f_{\rmn{v}0})/(9f_{\rmn{v}0})$. In
Fig.~\ref{fig:hod}, $D=d_L/(1+z)$, and the dressed Hubble constant $H_0$, is
related to the bare Hubble constant $\bar{H}_0$ by
\be\label{eq:dressedH0}
H_0=(4f_{\rmn{v}0}^2+f_{\rmn{v}0}+4)/[2(2+f_{\rmn{v}0})]\bar{H}_0.
\ee

The dressed Hubble constant is the one whose value should coincide with
that which is conventionally determined on scales greater than the scale
of statistical homogeneity, $z\gtrsim0.033$. Using the ``Gold'' \sn\ dataset of
\citet{gold}, \citet{LNW} found the dressed Hubble constant to be
$H_0= 61.7^{+1.4}_{-1.3}\kmsmpc$, (and the bare Hubble constant $\bar
H_0=48.2\pm2.6\kmsmpc$). However, since the supernova data magnitudes depend on
an overall normalization determined from the local distance ladder, they cannot
be used to determine the Hubble constant alone. Joint estimates of $H\Z0$ and
$\fvn$ which fit both the angular diameter distance of the sound horizon in the
CMB anisotropy data and the baryon acoustic oscillation scale in galaxy
clustering statistics do provide an independent constraint on the value of
$H\Z0$, however, and on this basis \citet{LNW} find the range of values
of the dressed Hubble constant to be roughly constrained to lie in
the interval $57\lsim H\Z0 \lsim 68\kmsmpc$.  This is lower than the S$H\Z0$ES
best estimate of \citet{shoes} but that survey relies on the calibration of the
distance ladder using objects that lie within the scale of statistical
homogeneity, and this may involve complicated systematics given that higher
values of $H\Z0$ are expected below the statistical homogeneity scale. Estimates
of $H\Z0$ which do not rely on calibration with nearby objects are often
somewhat lower. For example, \citet{courbin10} find $H_0= 62^{+6}_{-4}\kmsmpc$
using the time delay from strong gravitational lensing of quasars, and
\citet{beutler11} estimate $H_0=67\pm3.2\kmsmpc$ using the WMAP sound
horizon-calibrated BAO signal in the 6dF galaxy survey. These measurements
indicate that a value of $H_0$ consistent with the TS model is still to be
obtained once systematic errors on distance determinations are reduced.

With the tracker solution, the bare densities $\OMM$ and $\OMk$ can be written
in terms of the present void fraction. In particular, the dressed matter
density, measured by wall observers, for which the numerical value is most
likely to be similar to that of a FLRW model, will be
\be \label{eq:fv0Om}
\Omega_{\rmn{m}0} =
\frac{1}{8}(2+f_{\rmn{v}0})^3\OMM=\frac{1}{2}(1-f_{\rmn{v}0})(2+f_{\rmn{v}0})
\ee
for the tracker solution.

Since there is no nearby GRB sample, there is no GRB calibration of $\bar{H}_0$,
and we work with relative distances only. The fitting process therefore results
in a best-fit value for the single parameter $f_{\rmn{v}0}$.

\subsection{GRB data reduction method}
In this paper I will use the sample of 69 GRBs selected by \citet{s07}
(henceforward S07) as having sufficient light curve data to compute their
placement on a Hubble diagram. 

GRBs are not standard candles, since their luminosities span several orders of
magnitude (whether one assumes collimated or isotropic emission). However, there
are ongoing attempts to ``standardize'' GRBs given their promise for cosmology: 
they occur at higher redshifts than any established standard candles, and
radiation in the gamma band ($\geq10$ keV) is not subject to the same
limitations due to dust extinction as the optical band~\citep{ggf2006}. Certain
GRB light curve parameters have been found to correlate with each other,
offering the possibility of computing a magnitude, much in the same way as the
Phillips stretch-luminosity relation is used to reduce scatter in the \sn\
Hubble diagram. 

\citet{s07} uses four light curve parameters that correlate with the luminosity:
(1) the lag time $\lagt$ is the time shift between the hard and the soft light
curves; (2) the light curve variability $V$ is the normalized variance of the
light curve around a smoothed version of that light curve; (3) the peak energy
$\Epeak$ is the photon energy at which the $\nu F_{\nu}$ spectrum is brightest;
and (4) the minimum rise time $\trt$ is the shortest time over which the light
curve rises by half the peak flux of the pulse. {\bf A} fifth correlation,
$\Epeak-\Egamma$, relates the peak energy of the light curve to total photon
energy emitted by the burst. This is the tightest of the
correlations~\citep{g2004}, but it requires measurement of a jet break time
by which the measured (isotropic-equivalent) energy can be corrected for the
collimation. Along with these luminosity indicators a peak flux $P$ is also
measured for a wide range of bandpasses. A bolometric flux (or fluence)
can be calculated by extrapolating to high and low energies using the well-known
broken power law of \citet{band93} for the GRB spectrum and integrating over all
energies. This brings consistency to the brightnesses, and given a cosmological
model, permits the calculation of an isotropic luminosity
\be \label{eq:dl}
L=4\pi d_L^2\Pbolo.
\ee

The algorithm goes as follows. The luminosity indicator is the independent
variable, and from eq.~(\ref{eq:dl}) is obtained the $Y$-coordinate. A linear
fit to the logarithms of these quantities gives an empirical relationship
between the luminosity indicator and the luminosity. For this fit, we use the
bisector of the two ordinary least squares fits: that of $X$ against $Y$,
and then vice versa~\citep{isobe90}. We can then use this relationship to
calculate a theoretical luminosity curve for each indicator, based on the
luminosity distance obtained from eq.~(\ref{eq:dl}). In cases where a jet break
has been measured, $\Epeak$ is related to the collimation-corrected energy
$\Egamma$, by 
\be \label{eq:dl2}
\Egamma=4\pi d_L^2\Sbolo(1-\cos\thetajet)(1+z)^{-1},
\ee
for jet opening angle $\thetajet$ and bolometric fluence $\Sbolo$.
The uncertainties in the $Y$-axis quantities $\log~L$ and $\log~\Egamma$
are obtained from the uncertainties in the $X$-axis quantities in the standard
way. Because the physics of the GRB explosions is not completely understood, the
correlations contain some scatter over and above the measurement noise. To
account for this, an additional intrinsic uncertainty is estimated such that the
reduced $\chisq$ of the indicator-luminosity calibration curve is unity. The
best-fit lines for these relations are given along with their uncertainties in
Appendix~\ref{app:1}.

From each calculated $L$ or $\Egamma$ we then recalculate a luminosity
distance via (\ref{eq:dl}) or (\ref{eq:dl2}), from which we obtain a distance
modulus in the standard way: $\mu=5\log d_L - 25$ for $d_L$ in Mpc. The
propagated uncertainties are \citep{s07}
\be \label{eq:pbolounc}
\sigma_{\mu}^2=(2.5\sigma_{\log
L})^2+\Big(\frac{1.086\sigma_{\Pbolo}}{\Pbolo}\Big)^2,
\ee
if the bolometric flux $\Pbolo$ is used, or, if the bolometric fluence is used,
\be \label{eq:sbolounc}
\sigma_{\mu}^2=(2.5\sigma_{\log
\Egamma})^2+\Big(\frac{1.086\sigma_{\Sbolo}}{\Sbolo}\Big)^2+\Big(
\frac{1.086\sigma_{\Fbeam}}{\Fbeam}\Big)^2,
\ee
where the beam factor $\Fbeam\equiv(1-\cos\thetajet)$ is calculated from the
jet break time $\tjet$.

Finally, we take a weighted average of all the five different distance moduli:
\be \label{eq:muav}
\mu=\frac{\Sigma_i\mu_i/\sigma_{\mu_i}^2}{\Sigma_i\sigma_{\mu_i}^{-2}},
\ee
with the uncertainty
\be \label{eq:muaverr}
\sigma_{\mu}=\Big(\Sigma_i\sigma_{\mu_i}^{-2}\Big)^{-1/2}.
\ee

We avoid circularity by performing a simultaneous fit of both the cosmology and
the luminosity relations~\citep{g2004,s07}--- i.e. the luminosity relations are
part of the model. The value of $H_0$ here is arbitrary, since its variation
changes the Hubble line and the luminosity calibration of the data in an
identical way, resulting merely in a change in the overall normalization of the
Hubble diagram. GRBs do not occur in the local universe, so calibrating the GRB
Hubble diagram to a value of $H_0$ with any accuracy is not possible.
This is different to the \sn\ case, in which the calibration of light
curve and stretch can be done model-independently with nearby \sne, and then
extrapolated to objects at higher redshifts. However, regardless of the
normalization, the \emph{shape} of the curve in the Hubble diagram depends
solely on $\fvn$. This means that for a range of values of $\fvn$, here between
0.0 and 1.0, we calibrate the luminosity relation and compute the placement of
the GRBs on the Hubble diagram, and calculate a corresponding range of $\chisq$
values. The favoured value for $\fvn$ is that for which the $\chisq$ is
minimized.

\section{Results}\label{results}
\begin{figure}
\begin{center}
\caption{Power law relation between lag time $\lagt$, corrected to the GRB rest 
frame, and isotropic luminosity, for 38 GRBs. The $1\sigma$ measurement
uncertainties are used for the error bars. The \lcdm\ fit as calculated in
\citet{s07} is shown in grey. The intercept $a$ and the slope $b$ for the TS
calibration are shown on the plot, and the equation of the best-fit line and
the expected uncertainty in the luminosity so calculated is given in
Appendix~\ref{app:1}\ref{tlag}.
\label{fig:bfits1}}
\includegraphics[scale=0.7]{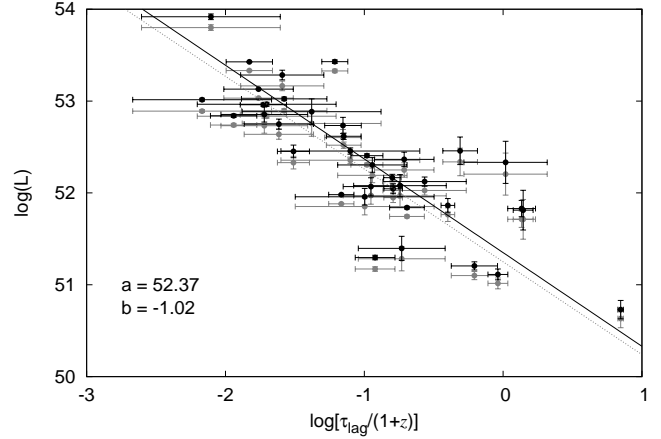}
\end{center}
\end{figure}

\begin{figure}
\begin{center}
\caption{Bisector fit of the Variability-Luminosity relation for 51 GRBs. Larger 
measurement uncertainties in this relation mean it carries less weight in
the final luminosity average. The TS intercept $a$ and the slope $b$
are shown, and the equation of the best-fit line and the expected
uncertainty in the luminosity so calculated are given in
Appendix~\ref{app:1}\ref{vl}.
\label{fig:bfits2}}
\includegraphics[scale=0.7]{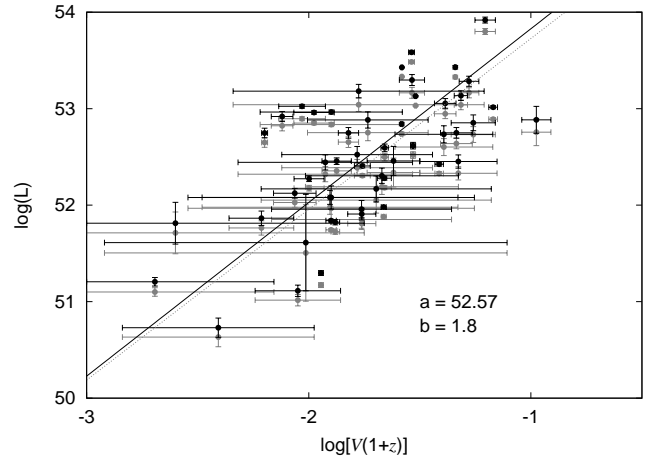}
\end{center}
\end{figure}

\begin{figure}
\begin{center}
\caption{Bisector fit of GRB isotropic luminosity to the peak energy $\Epeak$, 
corrected to the rest frame of the GRB. $N=64$. The TS intercept $a$ and the
slope $b$, and the equation of the best-fit line and the expected uncertainty in
the luminosity so calculated are given in Appendix~\ref{app:1}\ref{epl}.
\label{fig:bfits3}}
\includegraphics[scale=0.7]{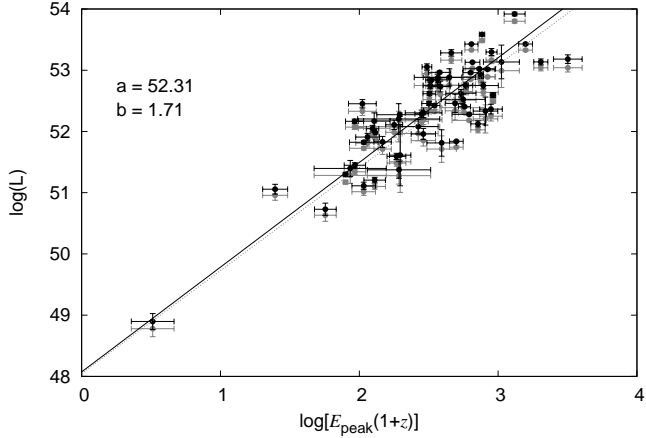}
\end{center}
\end{figure}

\begin{figure}
\begin{center}
\caption{$\Epeak-\Egamma$ relation for 27 GRBs. This is the tightest of the five 
power relations but there are fewer data points, since calculation of
$\Egamma$ requires identification and measurement of a jet break. The TS
intercept $a$ and the slope $b$ are shown on the plot, and the equation of the
best-fit line and the expected uncertainty in the $\Egamma$ so calculated are
given in Appendix~\ref{app:1}\ref{epeg}.\label{fig:bfits4}}
\includegraphics[scale=0.7]{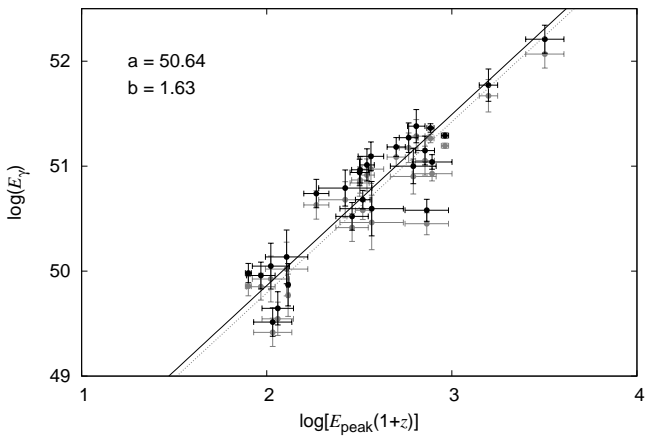}
\end{center}
\end{figure}

\begin{figure}
\begin{center}
\caption{Minimum rise time-Luminosity relation for 62 GRBs. The TS intercept $a$
and the slope $b$ are shown, and the equation of the best-fit line and the
expected uncertainty in the luminosity so calculated are given in
Appendix~\ref{app:1}\ref{trt}.
\label{fig:bfits5}}
\includegraphics[scale=0.7]{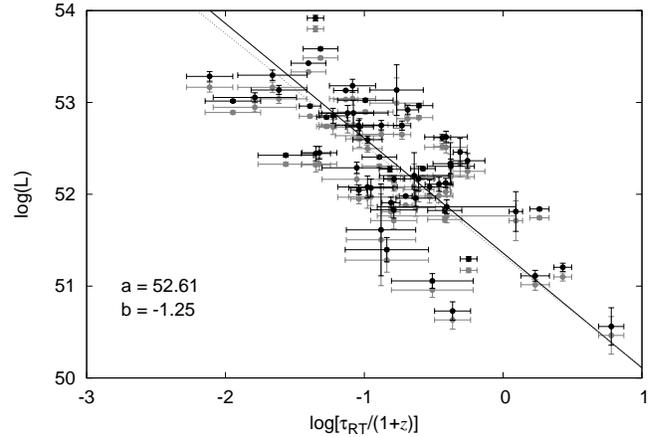}
\end{center}
\end{figure}

We show the results of the linear regressions in figures~\ref{fig:bfits1}--\ref
{fig:bfits5} in black. The intercept $a$ and slope $b$ of the TS model
calibration line are shown in each figure. For comparison, the \lcdm-calibrated
data points (for a flat Friedmann model with $\OmMn=0.27$ and $w=-1$, as
calculated in S07) are shown in grey.

The timescape model produces calibrations that are within 1$\sigma$ of the \lcdm\ 
model in each case. In fact, the TS model regression parameters match those of
the concordance model more closely than regression parameters calculated from
the variable dark energy equation of state cosmology of \citet{Riess2004}
($w_0=-1.31$, $w'=1.48$), computed in S07 to assess the dependence of the
calibration on the input cosmology.

\begin{figure}
\begin{center}
\caption{Hubble diagram for the 69 GRBs of S07. The \lcdm\ diagram is shown in
grey, and the TS diagram is shown in black. 
\label{fig:tshd}}
\includegraphics[scale=0.7]{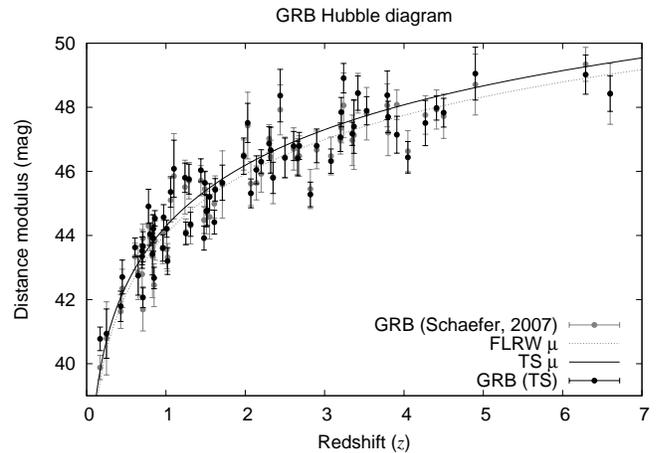}
\end{center}
\end{figure}

The resulting Hubble diagram for the TS model is shown in fig.~\ref{fig:tshd}.
In the \lcdm\ case, with the ``concordance'' value $\OmMn=0.27$, we obtain a
reduced $\chisq$ of 1.05 as in S07. The parameter values that minimize the HD
$\chisq$ are $\OmMn=0.21^{+0.22}_{-0.11}$ for the \lcdm\ model\footnote{This
coincides within a standard deviation with $\OmMn=0.39^{+0.12}_{-0.08}$ found in
S07, which was found by marginalizing over the slopes and intercepts of the
luminosity relations.}, shown in grey, and $\fvn=0.84^{+0.14}_{-0.21}$ for the
timescape model, for which the reduced $\chisq$ was 1.04 for 68 dof, shown in
black. This present void fraction corresponds to a matter density as measured by
wall observers via eq.~(\ref{eq:fv0Om}) of $\OmMn=0.23^{+0.25}_{-0.20}$.
However, note that there is no \emph{a priori} reason why the \lcdm\ and TS
values for $\OmMn$ should coincide, since the role of this parameter in each
theory is different.

The TS model fits the GRB Hubble diagram slightly better (lower $\chi^2$) than
the \lcdm\ model. The corresponding Bayes factor $\ln B=0.18$ indicates Bayesian
evidence in favour of the timescape model that is ``not worth more than a bare
mention" according to the Jeffreys scale \citep{kr1995}. This is apparent, since
the competing predictions of the models lie well within the range spanned by
the measurement errors, let alone the systematics, so it can only be concluded
that GRB cosmology is not yet precise enough to distinguish between these
models.

By contrast, preliminary investigations by \citet{schaeferTalk} indicate that
certain modified gravity models and particular exotic forms of dark energy (the
Chaplygin gas) provide much poorer fits to the GRB data than the standard \lcdm\
model. Amongst the alternatives to the \lcdm\ model, the TS model therefore
enjoys a degree of phenomenological success which is hard to replicate in a
number of other scenarios.

\section{Conclusion}\label{discussion}
Some issues with the use of GRBs as standard candles for constraining
cosmological parameters are discussed by \citet{ggf2006}, \citet{g2009}, and
\citet{p2009}. In particular, the correlations between the isotropic
luminosities and the luminosity indicators are weak in a $\chisq$
sense---physical factors unaccounted for are producing large scatter. Strictly
speaking, the $\Epeak-\Egamma$ correlation is the only relationship with a
sufficiently low reduced $\chisq$ to admit cosmological parameter estimation,
albeit with the caveat that the measurement of the jet break time
assumes a particular fireball model~\citep{g2009}. \citet{p2009} point out that
the luminosity correlations are statistical in nature, rather than being, as
they should ideally, one-to-one relations between uncorrelated quantities. This
meant, for example, that the comparatively tight $\Epeak-\Egamma$ correlation
found by \citet{g2004} actually weakened with the introduction of more data
points. 

Systematic uncertainties such as dust extinction and evolution constitute
considerable limitations to cosmological parameter estimation with
\sne. Many of these uncertainties, for example Malmquist bias and
gravitational lensing, are considered negligible in the redshift range over
which \sne\ occur. For the redshift range over which GRBs occur, one would
expect that Malmquist bias and lensing might cause at least some of the scatter
in the GRB Hubble diagram, but these biases are shown in S07 to be negligibly
small. Obscuration by dust is not an issue for GRBs~\citep{g2004}, but
selection and evolution effects can potentially influence the current GRB
sample. The well-known ``Amati'' correlation for long-duration GRBs between
isotropic-equivalent radiated energy $\Eiso$, describing the intensity of the
burst, and the photon energy at which the time-averaged spectrum peaks $\Epi$,
although proving to be quite robust~\citep*{amati08,amati10}, has shown
evidence of variation with redshift~\citep{li07} and susceptibility to
detector threshold selection effects~\citep{butler2007}. \citet{p2009}
find evidence for evolution of the GRB peak luminosities, but this should not
affect the Hubble diagram, since it is the luminosity relations which
should give the right distances for placement on the Hubble diagram.

It can be argued that the kind of relativistic and geometric effects that
underlie the luminosity relations should not be greatly affected by evolution or
the metallicity of the progenitor~\citep{s07}. It is conceivable that a better
understanding of GRB physics in the future will allow them to be used as
``standardizable'' candles, and put their utility for cosmological parameter
estimation and discrimination between cosmological models on a firmer basis.
Ongoing observational programmes such as \emph{Swift} continue to contribute to
this aim. We need to know more about the physics of the GRBs, and we need more
high-quality measurements of GRB redshifts, light curves and spectra.

In the meantime, however, we can obtain glimpses of the potential applications
of standard candles whose range extends into the era of decelerating cosmic
expansion. In the present study, the correlations are forced to be a good fit by
incorporating the additional ``systematic error'' term, computed such that it
makes the $\chisq$ of the best-fit correlation equal to one. This term
contributes (in quadrature) to the uncertainty in the log of the isotropic
luminosity which propagates through to the Hubble diagram $\chisq$ via eqs
(\ref{eq:pbolounc}) and (\ref{eq:sbolounc}). A single GRB at a redshift of 5 or
6, with better-determined physical characteristics, potentially carries
more statistical power than a single \sn\ at $z=1.7$ because of the Hubble
diagram ``lever arm''---the Hubble diagram at redshifts $z>2$ changes
with a different cosmological model or cosmological parameters much more than it
does at lower redshifts. We obtain results that are not inconsistent with
current models, with certain acknowledged caveats. In particular, there is much
scope for progress in improving the GRB Hubble diagram, and much to be gained.

\section*{Acknowledgments}

I thank David Wiltshire for suggestions and discussions. 
This work was supported by a University of Canterbury Doctoral Scholarship
and the Marsden fund of the Royal Society of New Zealand.

\appendix
\section{Timescape calibration curve equations}\label{app:1}

The five calibrations in figures~\ref{fig:bfits1}--\ref{fig:bfits5} are based
on the bisector of the two ordinary least squares fits~\citep{isobe90}. For the
$i$th luminosity indicator, the best-fit line has the form $Y_i=a+bX_i$, where
$X_i=\log($indicator$)\pm\log(1+z)$, where the sign of the redshift factor
depends on the indicator. The five best-fit lines for the TS model (those for
which the HD $\chisq$ is a minimum), their associated uncertainties,
and the corresponding \lcdm\ values for $a$ and $b$, are given below.
\begin{enumerate}
\item\noindent\emph{Lag time vs. Luminosity}: \label{tlag}\\
For the timescape model:
\be
\log~L=52.37-1.02\log\Big[\frac{\lagt(1+z)^{-1}}{0.1~\rmn{s}}\Big];
\ee
\be
\sigma^2_{\log~L}=&\sigma_a^2& +~
\Big\{\sigma_b\log\Big[\frac{\lagt(1+z)^{-1}}{0.1~\rmn{s}} \Big ]\Big\}^2
\\\nonumber &+&\Big(\frac{0.4343b\sigma_{\rmn{lag}}}{\lagt}\Big)^2 +
\sigma^2_{\rmn{lag,sys}},
\ee
where $\sigma_a=0.13$, $\sigma_b=0.09$, and $\sigma^2_{\rmn{lag,sys}}=0.37$
gives a reduced $\chisq$ of one. For the \lcdm\ calibration, we find
$a=52.30\pm0.13$, $b=-1.00\pm0.09$ and $\sigma^2_{\rmn{lag,sys}}=0.36$.
\vspace*{0.5cm}
\item\noindent\emph{Variability vs. Luminosity}: \label{vl}\\
For the timescape model:
\be
\log~L=52.57+1.80\log\Big[\frac{V(1+z)}{0.02}\Big];
\ee
\be
\sigma^2_{\log~L}=&\sigma_a^2& +~
\Big\{\sigma_b\log\Big[\frac{V(1+z)}{0.02} \Big ]\Big\}^2
\\\nonumber &+&\Big(\frac{0.4343b\sigma_{V}}{V}\Big)^2 +
\sigma^2_{V\rmn{,sys}},
\ee
where $\sigma_a=0.34$, $\sigma_b=0.20$, and $\sigma^2_{V\rmn{,sys}}=0.35$
gives a reduced $\chisq$ of one. For the \lcdm\ calibration, we find
$a=52.50\pm0.34$, $b=1.77\pm0.20$ and $\sigma^2_{V\rmn{,sys}}=0.35$.
\vspace*{0.5cm}
\item\noindent\emph{$\Epeak$ vs. Luminosity}: \label{epl}\\
For the timescape model:
\be
\log~L=52.31+1.71\log\Big[\frac{\Epeak(1+z)}{300~\rmn{keV}}\Big];
\ee
\be
\sigma^2_{\log~L}=&\sigma_a^2& +~
\Big\{\sigma_b\log\Big[\frac{\Epeak(1+z)}{300~\rmn{keV}} \Big ]\Big\}^2
\\\nonumber &+&\Big(\frac{0.4343b\sigma_{\Epeak}}{\Epeak}\Big)^2 +
\sigma^2_{\Epeak\rmn{,sys}},
\ee
where $\sigma_a=0.24$, $\sigma_b=0.10$, and $\sigma^2_{\Epeak\rmn{,sys}}=0.34$
gives a reduced $\chisq$ of one. For the \lcdm\ calibration, we find
$a=52.24\pm0.24$, $b=1.69\pm0.10$ and $\sigma^2_{\Epeak\rmn{,sys}}=0.34$.
\vspace*{0.5cm}
\item\noindent\emph{$\Epeak$ vs. $\Egamma$}: \label{epeg}\\
For the timescape model:
\be
\log~\Egamma=50.64+1.63\log\Big[\frac{\Epeak(1+z)}{300~\rmn{keV}}\Big];
\ee
\be
\sigma^2_{\log~\Egamma}=&\sigma_a^2& +~
\Big\{\sigma_b\log\Big[\frac{\Epeak(1+z)}{300~\rmn{keV}} \Big ]\Big\}^2
\\\nonumber &+&\Big(\frac{0.4343b\sigma_{\Epeak}}{\Epeak}\Big)^2 +
\sigma^2_{\Egamma\rmn{,sys}},
\ee
where $\sigma_a=0.28$, $\sigma_b=0.10$, and $\sigma^2_{\Egamma\rmn{,sys}}=0.17$
gives a reduced $\chisq$ of one. For the \lcdm\ calibration, we find
$a=50.58\pm0.28$, $b=1.62\pm0.10$ and $\sigma^2_{\Egamma\rmn{,sys}}=0.15$.
\vspace*{0.5cm}
\item\noindent\emph{Rise time vs. Luminosity}: \label{trt}\\
For the timescape model:
\be
\log~L=52.61-1.25\log\Big[\frac{\trt(1+z)^{-1}}{0.1~\rmn{s}}\Big];
\ee
\be
\sigma^2_{\log~L}=&\sigma_a^2& +~
\Big\{\sigma_b\log\Big[\frac{\trt(1+z)^{-1}}{0.1~\rmn{s}} \Big ]\Big\}^2
\\\nonumber &+&\Big(\frac{0.4343b\sigma_{\Epeak}}{\Epeak}\Big)^2 +
\sigma^2_{\trt\rmn{,sys}},
\ee
where $\sigma_a=0.11$, $\sigma_b=0.11$, and $\sigma^2_{\trt\rmn{,sys}}=0.48$
gives a reduced $\chisq$ of one. For the \lcdm\ calibration, we find
$a=52.54\pm0.11$, $b=-1.23\pm0.11$ and $\sigma^2_{\trt\rmn{,sys}}=0.47$.

\end{enumerate}

\end{document}